\documentclass[pra,amsmath,amssymb,twocolumn,superscriptaddress]{revtex4-2}
\usepackage{amsmath}
\usepackage{amssymb}
\usepackage{amstext}
\usepackage{amsfonts}
\usepackage{amsxtra}
\usepackage{bm}
\usepackage[usenames]{color}
\usepackage{grffile}
\usepackage{soul}
\usepackage{svg}
\newcommand{\rr}{\mathbf{r}}
\newcommand{\RR}{\mathbf{R}}

\newcommand{\epb}{\bm{\epsilon}}
\newcommand{\dx}{\text{d}^3\textbf{x}\,}
\newcommand{\edd}{\varepsilon_\text{dd}}

\begin{document}

\title{Maintaining supersolidity in one and two dimensions}

\author{E. Poli}
 \affiliation{
     Institut f\"{u}r Experimentalphysik, Universit\"{a}t Innsbruck, Austria
 }

\author{T. Bland}
 \affiliation{
     Institut f\"{u}r Quantenoptik und Quanteninformation, \"Osterreichische Akademie der Wissenschaften, Innsbruck, Austria
 }
 
\author{C. Politi}
 \affiliation{
     Institut f\"{u}r Quantenoptik und Quanteninformation, \"Osterreichische Akademie der Wissenschaften, Innsbruck, Austria
 }
 \affiliation{
     Institut f\"{u}r Experimentalphysik, Universit\"{a}t Innsbruck, Austria
 }
 
\author{L. Klaus}
 \affiliation{
     Institut f\"{u}r Quantenoptik und Quanteninformation, \"Osterreichische Akademie der Wissenschaften, Innsbruck, Austria
 }
 \affiliation{
     Institut f\"{u}r Experimentalphysik, Universit\"{a}t Innsbruck, Austria
 }

\author{M. A. Norcia}
 \affiliation{
     Institut f\"{u}r Quantenoptik und Quanteninformation, \"Osterreichische Akademie der Wissenschaften, Innsbruck, Austria
 }

\author{F. Ferlaino}
 \affiliation{
     Institut f\"{u}r Quantenoptik und Quanteninformation, \"Osterreichische Akademie der Wissenschaften, Innsbruck, Austria
 }
 \affiliation{
     Institut f\"{u}r Experimentalphysik, Universit\"{a}t Innsbruck, Austria
 }

\author{R. N. Bisset}
 \affiliation{
     Institut f\"{u}r Experimentalphysik, Universit\"{a}t Innsbruck, Austria
 }

\author{L. Santos}
 \affiliation{
     Institut f\"{u}r Theoretische Physik, Leibniz Universit\"{a}t Hannover, Germany
 }

\begin{abstract}

We theoretically investigate supersolidity in three-dimensional dipolar Bose-Einstein condensates.
We focus on the role of trap geometry in determining the dimensionality of the resulting droplet arrays, which range from one-dimensional to zigzag, through to two-dimensional supersolids in circular traps.
Supersolidity is well established in one-dimensional arrays, and may be just as favorable in two-dimensional arrays provided that one appropriately scales the atom number to the trap volume.
We develop a tractable variational model--which we benchmark against full numerical simulations--and use it to study droplet crystals and their excitations.
We also outline how exotic ring and stripe states may be created with experimentally-feasible parameters.
Our work paves the way for future studies of two-dimensional dipolar supersolids in realistic settings.

\end{abstract}
\date{\today}
\maketitle

%%%%%%%%%%%%%%%%%%%%%%%%%%%%%%%%%%%%%%%%%%
\section{Introduction}

A supersolid concurrently exhibits both superfluidity and crystalline order \cite{gross1957unified,andreev1969quantum,thouless1969flow,chester1970speculations,leggett1970can,boninsegni2012colloquium}.
Although predicted over half a century ago, supersolidity was only recently realized in experiments: a feat made possible by the flexibility and high-degree of control afforded by quantum gas systems.
While supersolid properties were observed in experiments with cavity-mediated interactions \cite{leonard2017supersolid} and spin-orbit coupling \cite{li2017stripe,Bersano2019}, those platforms produced rigid lattices that are impervious to the usual excitations expected of crystals.
In contrast, supersolids with deformable crystals have now been realized in dipolar Bose-Einstein condensates \cite{Tanzi2019,Bottcher2019,Chomaz2019}, in which genuine crystal and superfluid excitations have been observed \cite{natale2019excitation,tanzi2019supersolid,guo2019low}. 

Dipolar Bose-Einstein condensates (BECs) can be obtained from highly-magnetic atoms such as chromium \cite{Griesmaier2005a}, dysprosium \cite{Mingwu2011a} and erbium \cite{Aikawa2012a}.
It was already predicted in 2003 that dipolar BECs could undergo a roton instability \cite{Santos2003a}--where the unstable excitations occur at finite momenta--as observed in cigar-shaped Er BECs \cite{Chomaz2018a,natale2019excitation} and more recently in a pancake-shaped Dy BEC \cite{schmidt2021roton}.
However, it was also expected from theory that the ensuing periodic density modulations would undergo a runaway collapse, and the regions of high local density would invoke 3-body losses that rapidly destroy the underlying BEC.
Indeed, a similar process was observed with the implosion of entire chromium BECs, driven by the attractive head-to-tail dipolar interactions \cite{Lahaye2009a}.
From the perspective of supersolidity, the missing ingredient was a mechanism to stabilize against such implosions,
and the answer came from the experimental discovery of dipolar droplets in Dy \cite{Kadau2016a,Schmitt2016a} and Er \cite{Chomaz2016} BECs.
Intriguingly, the stabilization mechanism is well-described by including the leading-order effects of quantum fluctuations, resulting in a theory now known as the extended Gross-Pitaevskii equation (eGPE) \cite{FerrierBarbut2016,Chomaz2016,Wachtler2016a,Bisset2016}. These beyond-mean-field effects are especially important for the highly-magnetic Er and Dy atoms.
With this knowledge in hand, the first dipolar supersolids were created by crossing the roton instability from the BEC regime to the droplet array regime \cite{Tanzi2019,Bottcher2019,Chomaz2019}, or directly by evaporative cooling into the supersolid phase \cite{Chomaz2019}.
The supersolid ground state region exists close to this phase transition, where the droplets overlap enough for the superfluid to globally conduct throughout the crystal.

While almost all dipolar supersolids have been experimentally realized as one-dimensional (1D) droplet arrays--see, for example, Refs.~\cite{Tanzi2019,Bottcher2019,Chomaz2019,natale2019excitation,tanzi2019supersolid,guo2019low}--two recent experiments have for the first time created two-dimensional (2D) supersolids \cite{norcia2021two,bland2021two}, thus opening an exciting frontier.
An early theoretical study in 2D predicted a rich phase diagram determined by competing metastable crystal configurations \cite{Baillie2018}.
More recent works in 2D have predicted supersolid edge phases \cite{roccuzzo2021supersolid}; intriguing manifestations of quantum vortices and persistent currents \cite{gallemi2020quantized,roccuzzo2020rotating,tengstrand2021persistent,ancilotto2021vortex}; honeycomb supersolids \cite{zhang2019supersolidity}; as well as ring and stripe phases \cite{zhang2021phases,hertkorn2021pattern}.

Associated with the rich physics on offer, dipolar supersolids have a large number of control parameters, and their effects on the ground state phase diagram interplay in a complicated way.
Furthermore, the supersolid regime typically lies only within a small range of parameters, located between the ordinary unmodulated BEC and a crystal of isolated droplets.
It is therefore paramount to develop strategies for maintaining supersolidity while exploring phase space.
From a theoretical perspective, it is also necessary to develop tractable and accurate descriptions to supplement the computationally intensive eGPE.

In this work, we study supersolidity in three-dimensional (3D) dipolar BECs. We systematically explore 1D and 2D droplet arrays, identifying the crucial role that the {\it average 2D density} has on maintaining supersolidity for various trap geometries and atom numbers.
We implement an eGPE formalism--and develop a tractable variational model--to examine the phase diagram from linear supersolids in elongated traps to 2D supersolids in circular traps, passing through zigzag and multi-row elliptical phases along the way. We find that 2D supersolids may be just as favorable as their 1D counterparts, provided that one fixes the average 2D density.
Through increasing the average 2D density we show how to observe the exotic ring and stripe phases \cite{zhang2021phases,hertkorn2021pattern} with realistic experimental parameters. Finally, we extend our variational model to study 2D crystal excitations, and benchmark this against full numerical calculations. 

The paper is structured as follows. In section \ref{sec:formalism} we outline our system and the eGPE, while section~\ref{sec:2dss} introduces the concept of the average 2D density, and uses it to theoretically build a 1D-2D supersolid phase diagram. We also introduce our droplet crystal variational model. Section~\ref{sec:increase} examines increasing the average 2D density to access the exotic ring and stripe phases. In section \ref{sec:excitations} we present some exemplary 2D crystal excitations, before concluding with section~\ref{Sec:Conc}.

%%%%%%%%%%%%%%%%%%%%%%%%%%%%%% Formalism

\section{Formalism}\label{sec:formalism}

%%%%%%%%%% GPE
We consider 3D dipolar BECs under harmonic confinement and we use the eGPE, given by \cite{Wachtler2016a,Bisset2016,FerrierBarbut2016,Chomaz2016}
\begin{align}
    &i\hbar\frac{\partial\Psi(\textbf{x},t)}{\partial t} =  \bigg[-\frac{\hbar^2\nabla^2}{2m}+\frac12m\left(\omega_x^2x^2+\omega_y^2y^2+\omega_z^2z^2\right)  \nonumber\\
    &+ \int\text{d}^3\textbf{x}'\, U(\textbf{x}-\textbf{x}')|\Psi(\textbf{x}',t)|^2  +\gamma_\text{QF}|\Psi(\textbf{x},t)|^3\bigg]\Psi(\textbf{x},t)\,,
    \label{eqn:GPE}
\end{align}
where $m$ is the mass and $\omega_i=2\pi f_i$ are the harmonic trap frequencies.
The wavefunction $\Psi$ is normalized to the total atom number $N=\int {\rm d}^3\mathbf{x}|\Psi|^2$.
For dilute gases, two-body interactions are well-described by the pseudo-potential,
\begin{align}
    U(\textbf{r}) = \frac{4\pi\hbar^2a_{\rm s}}{m}\delta(\textbf{r})+\frac{3\hbar^2a_\text{dd}}{m}\frac{1-3\cos^2\theta}{r^3}\,,
\end{align}
with the first term describing the short-range interactions governed by the s-wave scattering length $a_s$. The second term represents the anisotropic and long-ranged dipole-dipole interactions, characterized by dipole length $a_\text{dd}=\mu_0\mu_m^2m/12\pi\hbar^2$, with magnetic moment $\mu_m$ and vacuum permeability $\mu_0$.
We take the dipoles to be polarized along $z$, and $\theta$ is the angle between the polarization axis and the vector pointing from one of the interacting particles to the other.
We always consider $^{164}$Dy, such that $a_\text{dd}=130.8a_0$, \textcolor{black}{where $a_0$ is the Bohr radius}.
The final term in (\ref{eqn:GPE}) is the dipolar Lee-Huang-Yang correction arising from quantum fluctuations \cite{Lima2011a}, having coefficient
\begin{align}
    \gamma_\text{QF}=\frac{128\hbar^2}{3m}\sqrt{\pi a_s^5}\,\text{Re}\left\{ \mathcal{Q}_5(\edd) \right\} \, ,
\end{align}
with \textcolor{black}{$\mathcal{Q}_5(\edd)=\int_0^1 \text{d}u\,(1-\edd+3u^2\edd)^{5/2}$} being the auxiliary function, and the relative dipole strength is given by $\edd = a_{\rm dd}/a_{\rm s}$.
Note that $\mathcal{Q}_5$ can be calculated analytically (Appendix \ref{Sec:AppenSD}), but this is just a monotonically increasing function that is of order unity for the regimes that we consider here. Ground state and metastable solutions of Eq.~\eqref{eqn:GPE} are calculated by minimizing the energy functional corresponding to the eGPE using a conjugate-gradients technique \cite{Ronen2006a}.

\begin{figure}
    \centering
	\includegraphics[width=1\columnwidth]{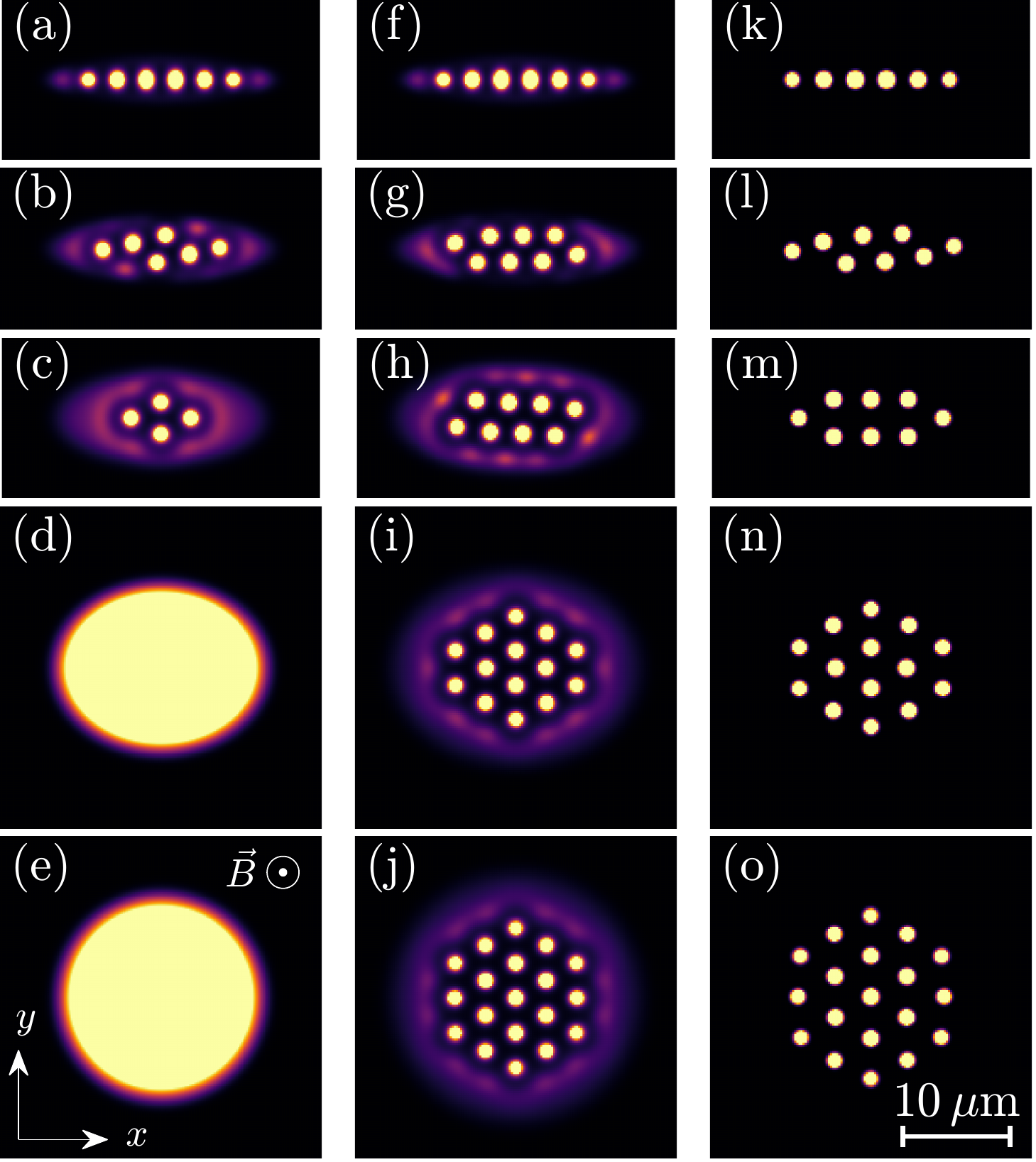}
\caption {
Opening up the trap from 1D to 2D for ${}^{164}$Dy atoms with $a_s=88a_0$ and $a_\text{dd}=130.8a_0$. In each panel we fix $(f_x,f_z)=(33,167)$\,Hz and decrease $f_y\in\{110,84.6,60,40,33\}$Hz, from top to bottom, showing the integrated column density. Column 1: eGPE result with constant $N=6.3\times10^4$. Column 2: eGPE with constant average 2D density, increasing $N$ to fix $\varrho  = Nf_xf_y$ with $N\in\{6.3, 8.19, 11.55, 17.325,  21\}\times10^4$. Column 3: same as Column 2 but the variational model. The atom number in the variational model is chosen to match the droplet atom number of the eGPE (see text). We always take the dipoles to be polarized by magnetic field $\vec B$ along $z$.} 
	 \label{fig:1} 
\end{figure}

%%%%%%%%%%%%%%%%%%%%%%%%%%%%%% Average 2D density
\section{Two-dimensional supersolidity}\label{sec:2dss}

%%%%%%%%%% Average 2D density
\subsection{Average 2D density}

In dipolar gases, the strong interplay between the confinement geometry and the long-ranged and anisotropic dipole-dipole interactions means that the ground state phase diagram is complex, and the relevant parameter space to consider is huge. This may conceal the identification of the most important control parameters.
For example, it was demonstrated in Refs.~\cite{Tanzi2019,Bottcher2019,Chomaz2019,Baillie2018,Roccuzzo2018,hertkorn2021pattern} that varying $a_s$ and $f_z$ dramatically affects the supersolid ground state, with supersolidity easily being lost.
In what follows we identify an important control parameter for moving between or within the various supersolid regimes, as well as maintaining supersolidity while progressing from 1D to 2D droplet arrays.

Dipolar supersolids require tight confinement along the direction of dipole polarization, and the precise choice of $f_z$ determines the narrow range of $a_s$ over which supersolidity occurs. For this reason, we take both $f_z$ and $a_s$ to be fixed in the following argument.
We propose that the \emph{average 2D density} acts as an important control parameter. This can be thought of as an \emph{average} over the droplet and interdroplet regions, and only the \emph{2D density} is considered because $f_z$ is fixed.
A simple yet powerful estimate for how the average 2D density scales is furnished by the Thomas-Fermi approximation, where kinetic energy is neglected, and the $x$ and $y$ radii of a BEC scale $\sim 1/f_x$ and $\sim 1/f_y$, respectively, giving a BEC area scaling $\sim 1/f_xf_y$.
The key point is then to realize that the average 2D density scales approximately with the parameter $\varrho = Nf_xf_y$.
In the next section, we explore the consequences of varying $\varrho$, versus keeping it fixed.

%%%%%%%%%%%%%%%%%%%%%%%%
\subsection{From 1D to 2D}\label{sec:1Dto2D}

In order to illustrate the utility of the average 2D density--characterized by $\varrho$--the first two columns of Fig.~\ref{fig:1} explore the 1D-2D transition for two different phase-space trajectories: first by allowing $\varrho$ to vary, and second by fixing $\varrho$.
For both, we consider fixed interactions while moving from a cigar-shaped trap (top row) to a pancake-shaped trap (bottom row).
The key difference between the trajectories is that column 1 has a fixed atom number--hence $\varrho$ decreases as the trap loosens--while column 2 instead fixes $\varrho$, with $N$ increasing to compensate for the widening of the trap.
Crucially, the reduction of $\varrho$ in the first column leads to a loss of the supersolid phase, replaced by an unmodulated BEC, while fixing $\varrho$ allows us to loosen the trap while remaining in the supersolid regime, eventually resulting in a large, 19-droplet supersolid for the circular trap [Fig.~\ref{fig:1}(j)].
We have theoretically verified in other work that this large 2D supersolid state is robust against thermal fluctuations \cite{bland2021two}.

%%%%%%%%%%%%%%%%%%%%%%%% Variational model formalism
\subsection{Droplet variational theory}

Although direct simulations of the eGPE have a remarkable predictive power, they are numerically intensive, and hinder a thorough overview.
We develop a variational model that permits a much simpler determination of the droplet phases available, while presenting an excellent qualitative, and largely quantitative, agreement with our eGPE calculations.

Inspired by recent work with nondipolar droplets \cite{Lavoine2021}, we assume the following ansatz for a dipolar droplet,
\begin{align}
\Psi(\textbf{x})=\sqrt{\mathcal{N}}\phi(\rho)\psi(z)\,,
\label{eqn:var}
\end{align}
with $\mathcal{N}$ the number of particles and $\rho=\sqrt{x^2+y^2}$. We again consider dipoles polarized along the $z$-axis, and the droplets are cylindrically symmetric, which we have confirmed as a good approximation by comparing with full eGPE calculations.
The radial and axial functions take the form, respectively:
\begin{align}
\begin{aligned}
\phi(\rho)&=\sqrt{\frac{r_\rho}{2\pi\Gamma(2/r_\rho)\sigma_\rho^2}}e^{-\frac{1}{2}\left( \frac{\rho}{\sigma_\rho}\right)^{r_\rho}}, \\
\psi(z)&=\sqrt{\frac{r_z}{2\Gamma(1/r_z)\sigma_z}}e^{-\frac{1}{2}\left( \frac{|z|}{\sigma_z}\right)^{r_z}}\,,
\label{eqn:var2}
\end{aligned}
\end{align}
with $\Gamma(x)$ being the Gamma function. The widths $\sigma_{\rho,z}$ and the exponents $r_{\rho,z}$ are variational parameters.
Note that this function permits the interpolation between a Gaussian ($r=2$) and a flat-top ($r\gg 1$) profile in a natural way. Furthermore, this ansatz allows for a simple evaluation of the various energies in the system using well-known properties of the Gamma function.

Our general strategy is to first numerically minimize the single-droplet problem for a range of possible parameters to build interpolation functions for the variational widths $\sigma_{\rho,z}(\mathcal{N})$ and exponents $r_{\rho,z}(\mathcal{N})$. These functions are then used to solve the many-droplet problem.

For a single droplet, ansatz (\ref{eqn:var}-\ref{eqn:var2}) can be used to minimize the eGPE energy functional, 
\begin{align}
    E_\text{sd}(\mathcal{N}) = E_\text{kin} + E_\text{trap} + E_\text{sr} + E_\text{dd} + E_\text{qf}\,,
    \label{eqn:energy}
\end{align}
where these quantities are the kinetic, trap, short-range interaction, dipole-dipole interaction, and quantum fluctuation contributions, respectively. The evaluation of these terms is detailed in Appendix \ref{Sec:AppenSD}.

Now consider a droplet array with $N_{\rm D}$ droplets, with $N_j$ atoms in the $j$-th droplet. 
Within the variational model, the energy of the droplet array is then given by:
\begin{align}
E = \sum_{j=1}^{N_{\rm D}} \left [  E_\text{sd}(N_j) + \frac{m}{2}(\omega_x^2 x_j^2+\omega_y^2 y_j^2) N_j \right ] + \sum_{j=1}^{N_{\rm D}}\sum_{j'> j} E_{jj'}\,, 
\label{eqn:fullvar}
\end{align}
where $E_{jj'}$ is the inter-droplet interaction, detailed in Appendix \ref{Sec:AppenIDI}.
By solving the single- then multi-droplet problems separately, we effectively reduce the number of variational parameters from $7N_{\rm D}-1$ to $3N_{\rm D}-1$ ($\{\sigma^j_{\rho,z},\,r^j_{\rho,z},N_j,x_j,y_j\}\to \{N_j,x_j,y_j\}$), where the $-1$ arises from fixing the total atom number, ${N=\sum_j N_j}$.

It is worth noting that important early work employed a purely Gaussian variational model (i.e.~$r_\rho=r_z=2$) to explore crystal and supersolid configurations \cite{Baillie2018}. Our model goes a step further by allowing for the possibility of droplets with flat-top density profiles, which partially acts to shield inter-droplet repulsion in the supersolid regime where the droplets are tightly packed together. 

\begin{figure}
    \centering
	\includegraphics[width=\columnwidth]{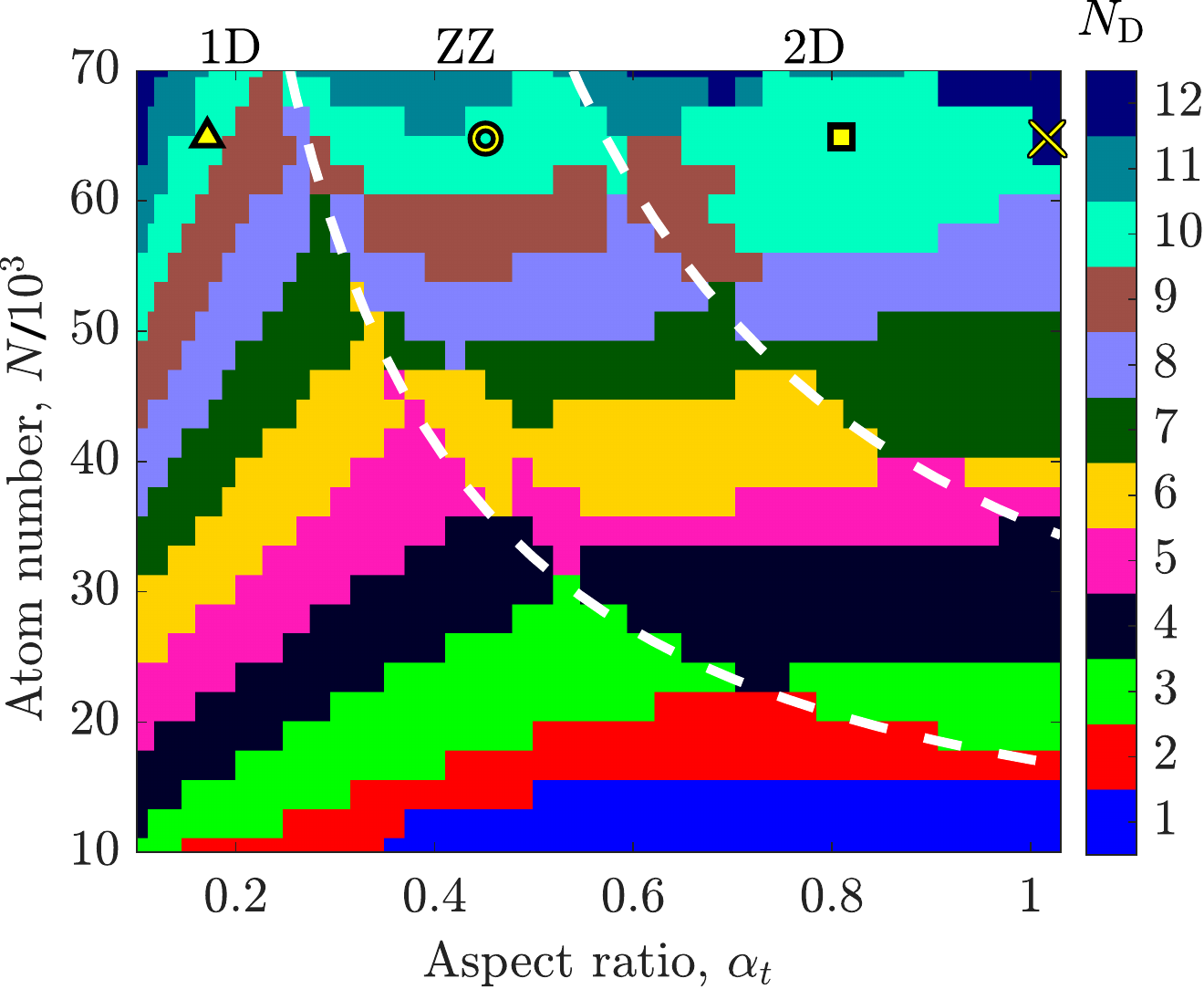}\\
		\includegraphics[width=\columnwidth]{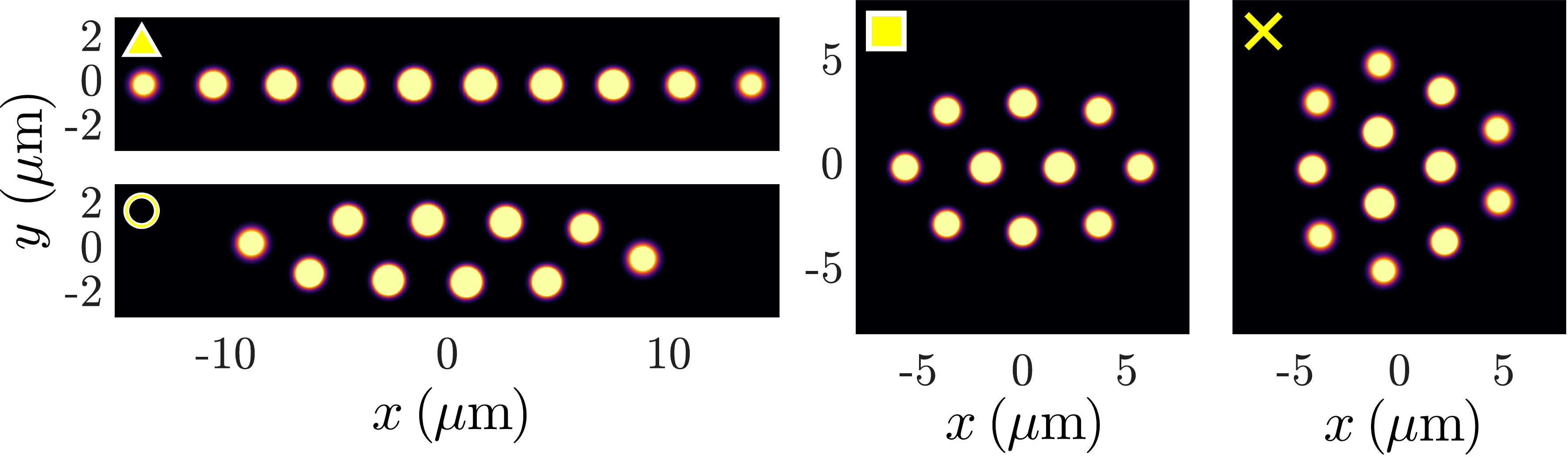}
\caption {Crystal phase diagram for $^{164}$Dy atoms from 1D (left) to circular trap regime (right) using ansatz (\ref{eqn:var}-\ref{eqn:var2}). Color indicates ground state droplet number versus total atom number $N$ and aspect ratio $\alpha_t=f_x/f_y$. A constant average 2D density (controlled by fixing $\varrho=Nf_xf_y$) is used throughout, which means the trap tightens from $\sqrt{f_xf_y}=43$\,Hz (top) to $\sqrt{f_xf_y}=114$\,Hz (bottom).
White lines separate the 1D, zigzag (ZZ) and 2D regions. Example configurations for fixed $N=5.4\times10^4$ are shown below. Parameters $f_z = 167$\,Hz and $a_s=88a_0$ remain constant.}
	 \label{fig:2} 
\end{figure}

Example solutions of our variational ansatz are shown in Fig.~\ref{fig:1}(column 3), displaying excellent agreement with the corresponding eGPE results (column 2).
It should be noted that for the eGPE solutions, a sizeable number of atoms exist outside the droplets in an outer ring, which we term the ``halo''.
To make direct comparisons between the variational and eGPE methods, we estimate the total number of atoms in the droplets alone from the eGPE and use this to set the total atom number for the corresponding variational calculation. For reference, the variational to eGPE atom number ratio varies from $N_\text{var} = 0.84N_\text{eGPE}$ for the linear chain [Figs.~\ref{fig:1}(f,k)] to $N_\text{var} = 0.58N_\text{eGPE}$ for the circular crystal [Figs.~\ref{fig:1}(j,o)].
Small deviations in the droplet positions occur between the models due to repulsion between the droplets and the halo in the eGPE, whereas the halo is absent in the variational model.
\textcolor{black}{In general, the halo leads to a slight compression of the crystal.  Additionally, because the halo density is nonuniform around the perimeter of the droplet array (in some cases forming nearly-droplet-like regions of higher density), its presence can also qualitatively modify the structure and the symmetry of the droplet array in certain situations [cf. Figs.~\ref{fig:1}(h,m)].}

%\textcolor{blue}{The halo density bulges in certain regions, acting like partially-formed droplets. In addition to a slight compression of the crystal, occasionally this can affect its symmetry too [cf.~Figs.~\ref{fig:1}(h,m)].}

%%%%%%%%%%%%%%%%%%%%%%%%%%%%%%
\subsection{Crystal phase diagram}

Here, with the variational model we seek to explore the full phase diagram of droplet crystal configurations whilst maintaining a fixed average 2D density, which we control by keeping $\varrho$ constant. 
Figure~\ref{fig:2} shows the droplet configurations of the ground state as a function of the trap aspect ratio $\alpha_t=f_x/f_y$ and atom number.
Since $\varrho$ is held fixed throughout, the bottom of the phase diagram corresponds to $N=10^4$ and $\sqrt{f_xf_y}=114$\,Hz, while the top reaches $N=7\times10^4$ and $\sqrt{f_xf_y}=43$\,Hz.
Traversing right on the phase diagram equates to increasing $f_x$ and decreasing $f_y$, hence moving to more circular configurations.

Several trends are apparent from this phase diagram. Larger $N$ corresponds to ground states with a larger number of droplets.
If the configuration is linear [left in Fig.~\ref{fig:2}], then the droplet number increases incrementally one droplet at a time, however for large $\alpha_t\sim1$ [right in Fig.~\ref{fig:2}] there are occasional jumps of two or more droplets--within the resolution of our phase diagram--corresponding to preferential triangular configurations of the lattice in 2D. For example, we find that for $\alpha_t=1$ the ground state jumps from $N_{\rm D}=8$ to the $N_{\rm D}=12$ state shown in Fig.~\ref{fig:2} $\times$, with only a very narrow range of $N$ corresponding to a 10 droplet configuration in between.

Following the solutions from bottom left to top right, in Fig.~\ref{fig:2}, there are two distinct jumps in the average transversal spread (${\Delta y=1/N_{\rm D}\sum_j^{N_{\rm D}}|y_j-\bar{y}|}$, for the $y$ position of the $j^\text{th}$ droplet $y_j$, and mean $y$ position $\bar{y}$), marked as white dashed lines on Fig.~\ref{fig:2}. These signify the transition from linear (Fig.~\ref{fig:2} $\bigtriangleup$) to zigzag (Fig.~\ref{fig:2} $\bigcirc$) configurations, and then 2D solutions with three (Fig.~\ref{fig:2} $\square$) or more (Fig.~\ref{fig:2} $\times$) rows of droplets. The first three of these highlighted solutions contain the same number of droplets for a fixed atom number, until $\alpha_t\approx1$ where the ground state configuration consists of 12 droplets. Intriguingly, these jumps in $\Delta y$ are also usually associated with a change in the ground state droplet number.
It is interesting to note that in the 1D regime, the regions of constant $N_{\rm D}$ slope downwards to the left.
This can be understood by considering a horizontal trajectory, for which both $N$ and $\varrho$ are constant. As we move left along this trajectory, increasing $f_y$ can no longer force the droplets closer together--since the array is already 1D--while the decreasing $f_x$ provides more space for longer droplet arrays, with larger $N_{\rm D}$.

%%%%%%%%%%%%%%%%%%%%% EXOTIC STATES

\section{Increasing average 2D density}\label{sec:increase}

\begin{figure}
    \centering
    \includegraphics[width=\columnwidth]{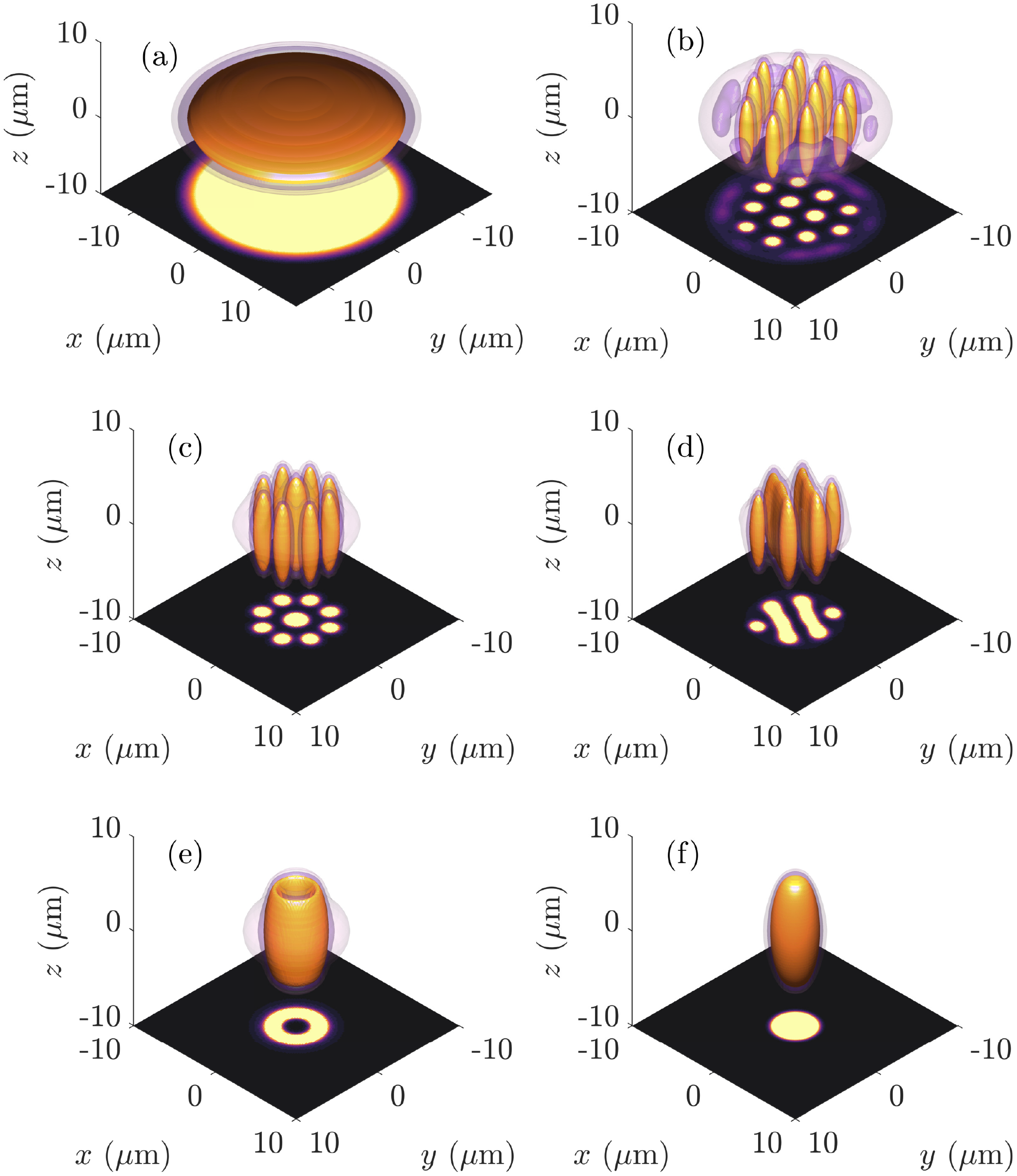}
    \caption{Increasing the average 2D density. The radial trap frequency is increased from (a)-(f), respectively, as $f_x=f_y\in\{30, 50, 80, 90, 100, 150\}$Hz, while $N=1.4\times10^5$ is held fixed. Density isosurfaces are shown at the 5\%, 0.1\%, and 0.01\% of the maximum density level. Shadow shows the 2D integrated density. Other parameters: $f_z=167$\,Hz, and $a_s = 88\,a_0$. }
    \label{fig:3}
\end{figure}

Previous theoretical works have found exotic two-dimensional supersolid states with either large atom numbers ($\sim10^6$), or tight trapping ($\sim1$kHz)  \cite{zhang2019supersolidity,zhang2021phases,hertkorn2021pattern}.
Notably, honeycomb ground states have been predicted \cite{zhang2019supersolidity} with crystal arrays of {\it holes} rather than droplets.
Such states are appealing due to their predicted strong superfluid conductance across the crystal, without relying on low density connections between droplets.
Also predicted are intriguing stripe and ring states \cite{zhang2021phases}, as well as labyrinthine instabilities \cite{hertkorn2021pattern} familiar in classical ferrofluids \cite{dickstein1993labyrinthine}.

Using the eGPE, we investigate the feasibility of creating these exotic supersolids by increasing the average 2D density through tightening the radial trap frequencies, without relying on pushing the parameters to unrealistically large values.
Figure \ref{fig:3}(a-f) shows how the solution changes by increasing $f_x=f_y\in\{30, 50, 80, 90, 100, 150\}$Hz, respectively, while holding fixed $N=1.4\times10^5$, hence $\varrho$ increases.
This trajectory through phase space takes us from an unmodulated BEC [Fig.~\ref{fig:3}(a)] to a hexagonal supersolid [Fig.~\ref{fig:3}(b)], a stripe supersolid [Fig.~\ref{fig:3}(d)], through to a ring state [Fig.~\ref{fig:3}(e)], and finally a macrodroplet [Fig.~\ref{fig:3}(f)].
Interestingly, while the peak density of the BEC phase is about $1.5\times 10^{20} \text{m}^{-3}$, for all droplet/supersolid phases it is roughly constant at $\sim1.5\times 10^{21} \text{m}^{-3}$, suggesting that the \textcolor{black}{atom losses from inelastic three-body collisions -- and hence also the lifetimes --} of these exotic states may be comparable to that for the current generation of supersolid experiments.

%%%%%%%%%%%%%%%%%%%%%%%%%%%%%%%%%%%% EXCITATIONS

\section{Excitations of a 2D supersolid}\label{sec:excitations}

\begin{figure}
    \centering
	\includegraphics[width=3.2in]{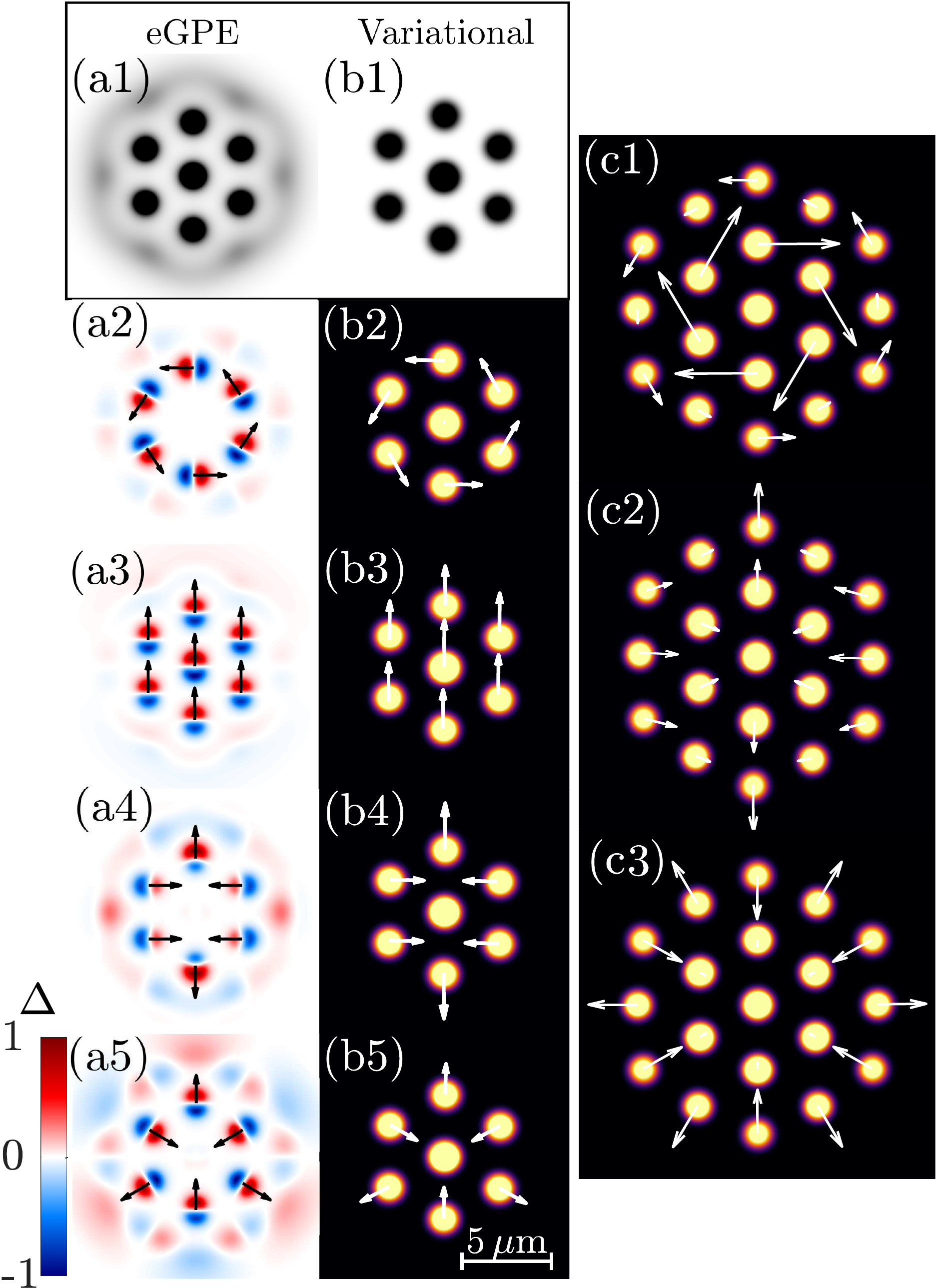}
\caption {
Crystal excitations.
(a1,b1) 7-droplet crystal state, and corresponding excitations from (a2-a5) eGPE-BdG calculations and (b2-b5) variational model.
Arrows indicate relative droplet motion \textcolor{black}{(see main text)}. Parameters: $a_s=90a_0$, $f_{x,y,z}=(52.83,52.83,167)$\,Hz, $N=9.5\times10^4$. (c) Exemplary excitations for the 19 droplet state from the variational model shown in Fig.~\ref{fig:1}(o).}
	 \label{fig:4} 
\end{figure}

Following the recent experimental observation of a 7-droplet hexagon supersolid \cite{bland2021two}, we further investigate the excitations of this state in a circular trap using the eGPE [see Fig.~\ref{fig:4}(a1)] and variational model [see Fig.~\ref{fig:4}(b1)].

We find excitations in the Bogoliubov-de Gennes (BdG) framework, which consists of a linearization of the eGPE around the stationary solution $\psi_0$ with perturbations of the form $\delta\psi = ue^{-i\epsilon t/\hbar}+v^*e^{i\epsilon t/\hbar}$ \cite{Pitaevskii16}.
To visualise the excitations we plot the density perturbation ${\Delta\psi = (u + v^*)|\psi_0|}$ for several exemplary excitations in Fig.~\ref{fig:4}(a2-a5) (arbitrary normalization).
\textcolor{black}{The arrows represent the droplet displacement vectors (with arbitrary global scaling), calculated from the shift in density peaks caused by adding a small amount of excitation to the ground state wavefunction.
These results are compared with the corresponding excitations calculated with the variational model [Fig.~\ref{fig:4}(b2-b5)], with droplet displacement vectors obtained through linearizing perturbations to the droplet positions [see Appendix \ref{Sec:AppenES}].}
Since these modes exist in the variational model--which does not account for superfluid flow between droplets--we can classify them as predominantly crystalline in nature.

Due to rotational symmetry there is a zero energy rotational mode [Figs.~\ref{fig:4}(a2,b2)], unique to circular trap supersolids.
As expected, there are two degenerate Kohn modes at the radial trap frequency, one of which is shown in Figs.~\ref{fig:4}(a3,b3).
Also plotted is a quadrupole excitations [Figs.~\ref{fig:4}(a4,b4)], as well as an example surface crystal mode [Figs.~\ref{fig:4}(a5,b5)], a unique feature of 2D supersolids highlighting the rich tapestry of excitations.
In the last two examples, the mode energy obtained in the BdG framework and the variational models differs. \textcolor{black}{The energies are $E/h=54$\,Hz [Fig.~\ref{fig:4}(a4)] and $E/h=72$\,Hz [Fig.~\ref{fig:4}(a5)] from the BdG calculations and $E/h=65$\,Hz [Fig.~\ref{fig:4}(b4)] and $E/h=69$\,Hz [Fig.~\ref{fig:4}(b5)] from the variational model.}
These deviations point to a measurable role played by the superfluid connection between the droplets, and the effect of the surrounding halo, which are not accounted for by the variational model.
Such comparisons between models provide an excellent platform to distinguish contributions from the crystal and the superfluid surrounding and connecting the droplets. 

The computational cost of obtaining modes from BdG linearization is high, requiring the diagonalization \textcolor{black}{of large dense matrices consisting of the total number of position space grid points squared, in our case $\sim 10^6\times 10^6$. We achieve this using an eigensolver based on the implicitly restarted Arnoldi method.}
We also find that the linearization is slower when there is no appreciable superfluid connection between the droplets, making excitations in the isolated droplet regime difficult to obtain.
\textcolor{black}{However, in this regime the variational model agrees well with the BdG calculations, and the former only requires the diagonalization of a $2N_\text{D} \times 2N_\text{D}$ matrix [i.e.~the total number of $(x_j,y_j)$ pairs]. This allows us to explore excitations of larger crystals.}

%However, similar crystal modes exist within the variational model, which only requires the diagonalization of a $(2N_\text{D})^2$ matrix (i.e.~the total number of $(x_j,y_j)$ pairs), and exact diagonalization is available for all relevant scattering lengths.

In Fig.~\ref{fig:4}(c1-c3) we show excitations of the 19 droplet crystal [Fig.~\ref{fig:1}(o)] using the variational model, a state that would require months of computational time to obtain excitations within the eGPE-BdG framework.
This configuration consists of two concentric hexagons with a single droplet in the middle, where the inner hexagon consists of 6 droplets and the outer hexagon has 12.
In Fig.~\ref{fig:4}(c1), we highlight an interesting mode in which the two outer hexagons counter-rotate.
We also find a quadrupole mode [Fig.~\ref{fig:4}(c2)], and in Fig.~\ref{fig:4}(c3) we show an analogue of the surface crystal mode we saw for the 7-droplet hexagon [Figs.~\ref{fig:4}(a5,b5)].

%%%%%%%%%%%%%%%%%%% CONCLUSIONS

\section{Conclusions}\label{Sec:Conc}

We have investigated the scope and feasibility of 2D supersolidity in harmonically trapped dipolar Bose gases, identifying the crucial role of the average 2D density in maintaining both the crystal structure and global superfluidity whilst varying the dimensionality and size of the droplet array.
By developing a variational multi-droplet model, we explored the phase diagram of crystal configurations for a wide range of atom numbers and aspect ratios for a fixed 2D density, identifying the transition from one- to two-dimensional droplet arrays.

We theoretically explored how increasing the average 2D density may provide a route for creating exotic stripe and ring supersolids under experimentally realistic conditions.
We also extended our variational model to explore crystal excitations, verified by direct comparison to the BdG analysis. This method allows for the investigation of crystal modes in large 2D supersolids, where exact diagonalization of the eGPE is demanding.

% Future work
Future work will further explore the potential of the variational model. Implementing a system of Hamilton equations would allow for dynamics of the droplet arrays, and further open up the study of excitations in two-dimensional supersolid crystals.
While we have revealed how to vary an important triplet of coupled parameters--$N$, and the two trapping frequencies perpendicular to the direction of dipole polarization, $f_x$ and $f_y$--enabling the exploration of supersolids of various shapes and sizes, future studies will seek an easy determination for how best to vary other control parameters, such as the coupling between the interaction strengths and the remaining trap frequency, $f_z$.

%%%%%%%%%%%%%%%%%%%%% ACKNOWLEDGEMENTS
\begin{acknowledgements}
We gratefully acknowledge useful discussions with Manfred Mark and the Innsbruck Erbium team. We acknowledge R.~M.~W.~van Bijnen for developing the code for our eGPE and BdG simulations. Part of the computational results presented have been achieved using the HPC infrastructure LEO of the University of Innsbruck. The experimental team is financially supported through an ERC Consolidator Grant (RARE, No.~681432), an NFRI grant (MIRARE, No.~OAW0600) of the Austrian Academy of Science, the QuantERA grant MAQS by the Austrian Science Fund FWF No I4391-N. L.S. and F.F.~acknowledge the DFG/FWF via FOR 2247/PI2790. L.S.~thanks the funding by the Deutsche Forschungsgemeinschaft (DFG, German Research Foundation) under Germany’s Excellence Strategy--EXC-2123 QuantumFrontiers--390837967. M.A.N.~has
received funding as an ESQ Postdoctoral Fellow from the European Union’s Horizon 2020 research and innovation programme under the Marie Sklodowska-Curie grant agreement No.~801110 and the Austrian Federal Ministry of Education, Science and Research (BMBWF). We also acknowledge the Innsbruck Laser Core Facility, financed
by the Austrian Federal Ministry of Science, Research and Economy.
\end{acknowledgements}

%%%%%%%%%%%%%%%%%%%%%%%% REFERENCES


%%%%%%%%%%%%%%%%%%%%%%%%% APPENDICES

\appendix

\section{Single-droplet variational model}\label{Sec:AppenSD}
Here we detail the individual contributions to the single-droplet energy functional for $\mathcal{N}$ atoms,
\begin{align}
    E_\text{sd}(\mathcal{N}) = E_\text{kin} + E_\text{trap} + E_\text{sr} + E_\text{dd} + E_\text{qf}\,. \label{Eq:EsdAppend}
\end{align}
These terms are given by
\begin{align}
\begin{aligned}
    E_\text{kin} &= -\frac{\hbar^2}{2m}\int \dx\, \Psi^*\nabla^2\Psi\,,\\
    E_\text{trap} &= \frac{m}{2}\int \dx\, \Psi^*\sum_i\omega_i^2x_i^2\Psi\,,\\
    E_\text{sr} &= \frac{1}{2}\int \dx\, \Psi^*g|\Psi|^2\Psi\,, \\
    E_\text{dd} &= \frac{g\epsilon_\text{dd}}{2}\int \frac{\text{d}^3\textbf{k}}{(2\pi)^3} \left ( \frac{3k_z^2}{k^2} -1 \right ) |\tilde n(\textbf{k})|^2\,, \\
    E_\text{qf} &= \frac25\gamma_{\text{QF}}\int\dx\,|\Psi|^5\,,
\end{aligned}
\label{eqn:energies}
\end{align}
 corresponding to the kinetic, trap, short-range interaction, dipole-dipole interaction, and quantum fluctuation contributions, respectively. Here, the short-range interaction coefficient is $g=4\pi\hbar^2a_s/m$, and the quantum fluctuation coefficient is  $\gamma_\text{QF}=\frac{32}{3}g\sqrt{\frac{a_s^3}{\pi}}\mathcal{Q}_5(\edd)$, where  \textcolor{black}{$\mathcal{Q}_5(\edd) = \text{Re}\left[\int_0^1 \text{d}u\, (1-\edd+3u^2\edd)^{5/2}\right]$,} and the density in Fourier space is $\tilde n(\textbf{k})=\int \dx\, e^{-i\textbf{k}\cdot\textbf{x}} |\Psi(\textbf{x})|^2$. The integral $\mathcal{Q}_5(\edd)$ can be evaluated as
\begin{widetext}
\begin{align}
    \mathcal{Q}_5(\edd) = \text{Re}
        \frac{(3\edd)^{5/2}}{48}\left[(8 + 26\epsilon + 33\epsilon^2)\sqrt{1 + \epsilon} + 15\epsilon^3\ln\left(\frac{1 + \sqrt{1 + \epsilon}}{\sqrt{\epsilon}}\right)\right]\,,
\end{align}
\end{widetext}
where $\epsilon = (1-\edd)/(3\edd)$.
Note, when using this definition care should be taken for the special cases $\mathcal{Q}_5(0)=1$ and $\mathcal{Q}_5(1)=3\sqrt{3}/2$.

These integrals are evaluated upon substitution of the ansatz $\Psi(\textbf{x})=\sqrt{\mathcal{N}}\phi(\rho)\psi(z)$ \cite{Lavoine2021} [see main text Eq.~\eqref{eqn:var}], with $\mathcal{N}$ the number of particles in the droplet. The radial and axial functions, normalized to one, are assumed to be of the form:
\begin{align}
\begin{aligned}
\phi(\rho)&=\sqrt{\frac{r_\rho}{2\pi\Gamma(2/r_\rho)\sigma_\rho^2}}e^{-\frac{1}{2}\left( \frac{\rho}{\sigma_\rho}\right)^{r_\rho}}, \\
\psi(z)&=\sqrt{\frac{r_z}{2\Gamma(1/r_z)\sigma_z}}e^{-\frac{1}{2}\left( \frac{|z|}{\sigma_z}\right)^{r_z}}\,,
\end{aligned}
\label{eqn:func}
\end{align}
with $\Gamma(x)$ the Gamma function. The widths $\sigma_{\rho,z}$ and the exponents $r_{\rho,z}$ are variational parameters. Substituting the ansatz Eqs.~\eqref{eqn:func} into the energy contributions Eqs.~\eqref{eqn:energies} gives the following results. 
The kinetic energy of the droplet is of the form: 
\begin{align}
\frac{E_\text{kin}}{\mathcal{N}}=\frac{\hbar^2}{2m\sigma_\rho^2}\frac{r_\rho^2}{4\Gamma(2/r_\rho)} 
+ \frac{\hbar^2}{2m\sigma_z^2}\frac{r_z f_K(r_z)}{2\Gamma(1/r_z)}\,, 
\end{align}
with $f_K(r_z)=(r_z-1)\Gamma(1-1/r_z) -\frac{r_z}{2}\Gamma(2-1/r_z)$.
The trap energy is:
\begin{align}
\frac{E_\text{trap}}{\mathcal{N}}=\frac{m}{2}\left ( \omega_x^2+\omega_y^2\right )\left [ \frac{\sigma_\rho^2\Gamma(4/r_\rho)}{2\Gamma(2/r_\rho)}\right ]
+\frac{m}{2}\omega_z^2 \left [ \frac{\sigma_z^2\Gamma(3/r_z)}{\Gamma(1/r_z)}\right ]\,.
\end{align}
Short-range interactions lead to an energy contribution:
\begin{align}
\frac{E_\text{sr}}{\mathcal{N}} = \frac{g\mathcal{N}}{8\pi \sigma_\rho^2\sigma_z}\frac{r_\rho r_z}{2^{2/r_\rho+1/r_z}\Gamma(2/r_\rho)\Gamma(1/r_z)}\,,
\end{align}
whereas quantum fluctuations result in the LHY correction:
\begin{align}
\frac{E_\text{qf}}{\mathcal{N}} =\frac{64 \mathcal{Q}_5(\epsilon_\text{dd})}{15\sqrt{\pi}}\left ( \frac{2}{5}\right )^{\frac{2}{r_\rho}+\frac{1}{r_z}}    gn_0 \sqrt{n_0 a^3}\,, 
\end{align}
where $n_0 = \frac{\mathcal{N} r_\rho r_z}{4\pi\Gamma(2/r_\rho)\Gamma(1/r_z)\sigma_\rho^2\sigma_z}$ is the central density.

The dipolar energy is best evaluated in momentum space. The ansatz density in Fourier space can be decomposed as $\tilde n(\textbf{k})=\tilde n_\rho(k_\rho)\tilde n_z(k_z)$, with
\begin{align}
\begin{aligned}
\tilde n_\rho(k_\rho)&=\frac{r_\rho}{\Gamma(2/r_\rho)}\int_0^\infty \text{d}\rho\, \rho e^{-\rho^{r_\rho}} J_0(k_\rho\sigma_\rho \rho)\,, \\
\tilde n_z(k_z) &= \frac{r_z}{\Gamma(1/r_z)} \int_0^\infty \text{d}z\, e^{-z^{r_z}}\cos(k_z\sigma_z z)\,,
\end{aligned}
\label{eqn:nk}
\end{align}
where $J_0$ is the first Bessel function of the first kind.

Interestingly, these functions can be very closely approximated by Gaussians:
$\tilde n_\rho(k_\rho)\simeq e^{-\alpha_\rho(r_\rho)(k_\rho\sigma_\rho)^2}$ and $\tilde n_z(k_z)\simeq e^{-\alpha_z(r_z)(k_z\sigma_z)^2}$, where $\alpha_\rho(\rho)$ and $\alpha_z(z)$ are functions found through numerical fitting to Eqs.~\eqref{eqn:nk} prior to variational minimization.
The DDI can be then easily expressed as
\begin{align}
\begin{aligned}
\frac{E_\text{dd}}{\mathcal{N}}=\frac{g\epsilon_\text{dd} \mathcal{N} f\left (\ell_\rho/\ell_z\right )}{2(2\pi)^{3/2}\ell_\rho^2\ell_z}\,,
\end{aligned}
\end{align}
where $\ell_{\rho,z}^2 = 4\alpha_{\rho,z}(r_{\rho,z})\sigma_{\rho,z}^2$, and 
\begin{align}
f(\kappa)=\frac{1}{\kappa^2-1}\left ( 2\kappa^2+1-3\kappa^2 \frac{\arctan(\sqrt{\kappa^2-1}) }{\sqrt{\kappa^2-1}}\right)\,.
\end{align}

Our approach is to first minimize the single droplet energy (\ref{Eq:EsdAppend}) for a suitable range of atom numbers.
Thus, in preparation for solving the multi-droplet problem, we generate interpolating functions $E_\text{sd}(\mathcal{N}),~ \sigma_{\rho,z}(\mathcal{N})$, and $r_{\rho,z}(\mathcal{N})$, furnishing a library of single-droplet solutions for a given trap and interaction parameters.

Employing this two-step method reduces the number of variational parameters from 7 per droplet to 3 ($\{\sigma^j_{\rho,z},\,r^j_{\rho,z},N_j,x_j,y_j\}\to \{N_j,x_j,y_j\}$).
Note that the final populations of the droplets are constrained by the total atom number $N=\sum_jN_j$.
The effect of inter-droplet repulsion is not accounted for in calculating the shape of the droplets. We replace $f_{x,y} \to 110$ Hz to simulate the effect of inter-droplet interactions on a given droplet’s shape, then to get the energy we use the $f_{x,y}$ of the actual trap. 

\textcolor{black}{For all minimization procedures related to variational calculations we use the sequential quadratic programming algorithm implemented in the MATLAB function \emph{fmincon}.}

\section{Inter-droplet interaction energy}\label{Sec:AppenIDI}

Let us consider two droplets with $N_1$ and $N_2$ atoms, respectively, which are sufficiently separated, such that we can neglect any overlapping.
The center-of-mass of the droplets is placed at $\textbf{r}_{j=1,2}=(x_j,y_j,0)$, i.e.~we permit displacements on the $xy$ plane, but assume that $z_j=0$. As for the single-droplet dipolar energy, the inter-droplet dipole-dipole interaction is best calculated in momentum space,
\begin{align}
E_{12}=g \epsilon_\text{dd} N_1 N_2 \int \frac{\text{d}^3\textbf{k}}{(2\pi)^3} \left [ 3\frac{k_z^2}{k^2}-1\right ] \tilde n_1^*(\textbf{k}) \tilde n_2(\textbf{k})\,, 
\end{align}
where we can approximate the Fourier transform of the density profile of the droplets as:
\begin{align}
\tilde n_j(\textbf{k}) \simeq e^{-k_\rho^2 \ell_\rho (N_j)^2/4}e^{-k_z^2\ell_z(N_j)^2/4} e^{ik_\rho(x_j\cos\phi + y_j\sin\phi)}\,.
\end{align}
The phase $\phi$ is accumulated due to the central position of the droplets being different from the origin, and plays no role in the energy calculation. We can then evaluate the interaction energy $E_{12}$, as a function of the distance $r_{12}=\sqrt{(x_1-x_2)^2+(y_1-y_2)^2}$ between the droplets:
\begin{align}
\!\!\!\! \!\!\!\! E_{12}(r_{12}) &= \frac{g\epsilon_\text{dd}N_1N_2}{\bar \ell_\rho^2 \bar \ell_z}\frac{\sqrt{2}}{\pi^2}  \nonumber \\
\!\!\!\!\!\!\!\!&\int_0^1 \text{d}u\, \frac{(\Lambda^2\!+\!2)u^2-\Lambda^2}{(1\!-\!\Lambda^2)u^2+\Lambda^2}
 G\!\left [ \frac{2r_{12}^2(1\!-\!u^2)}{\bar \ell_\rho^2}\right ]\,,
 \label{eqn:varint}
\end{align}
where $2\bar \ell_{\rho,z}^2=\ell_{\rho,z}(N_1)^2 + \ell_{\rho,z}(N_2)^2$, $\Lambda=\bar \ell_z/ \bar \ell_\rho$, and 
\begin{align}
G(x)=\frac{\sqrt{\pi}}{4} e^{-x/8} \left [ 
I_0\left(\frac{x}{8} \right)+\frac{x}{4}\left ( I_1\left(\frac{x}{8} \right )-I_0\left(\frac{x}{8} \right)\right )\right ]\,,
\end{align}
with $I_n(x)$ the modified Bessel function. The interaction energy \eqref{eqn:varint} is attractive at short distances, a spurious effect up to the radial size of a droplet. In order to prevent the droplets ``piling up” in this inner region, we instead approximate the inter-droplet potential as
\begin{align} 
E_{jj'}(r_{jj'})\simeq \frac{V_0(N_j,N_{j'}) N_j N_{j'}}{(r_{jj'}+r_0(N_j,N_{j'}))^3}\,,
\end{align}
for any two droplets $j$ and $j'$, where $V_0$ and $r_0$ are determined by fitting to equation Eq.~\eqref{eqn:varint}. This term is the last contribution to Eq.~\eqref{eqn:fullvar}, and is utilized in the phase diagram Fig.~\ref{fig:2}. By considering a range of particle number pairs between droplets we determine the interpolating functions $V_0(\mathcal{N},\mathcal{N}')$ and $r_0(\mathcal{N},\mathcal{N}')$ prior to solving the full many-droplet problem. Note that the shift $r_0$, which results from the $z$-extension of the droplet, is relevant because typical interdroplet distances are comparable to the $z$-size of the droplets.

\section{Excitations of the variational model}\label{Sec:AppenES}
Expanding around the equilibrium positions $\RR_j=(x_j,y_j)$, $\rr_j = \RR_j + \epb_j$, the energy of the array 
becomes, up to second order in the displacement $\epb_j=(\epsilon_{x;j}, \epsilon_{y;j})$, of the form $E = E_0 + E^{(2)}$ (the first order contribution cancels because we move from an energy minimum), with $E_0$ the ground-state energy, and
\begin{align}
E^{(2)} = \sum_{j=1}^{N_\text{D}} \epb_j^T \cdot \left [ \hat A_j \cdot \epb_j -\sum_{j'\neq j} \hat B_{jj'} \cdot \epb_{j'} \right ]\,,
\end{align}
where 
\begin{align}
\hat B_{jj'} &= V_0(N_j,N_{j'}) \sqrt{N_j N_{j'}}
\begin{pmatrix}
\beta_{jj'}  + \gamma_{jj'} X_{jj'}^2 & \gamma_{jj'}  X_{jj'} Y_{jj'}\\
\gamma_{jj'}  X_{jj'} Y_{jj'} & \beta_{jj'}  + \gamma_{jj'} Y_{jj'}^2 
\end{pmatrix}, \\
\hat A_j &= \frac{mN_j}{2}
\begin{pmatrix}
\omega_x^2 & 0\\
0 & \omega_y^2 
\end{pmatrix}
+\sum_{j'\neq j} \hat B_{jj'}\,,
\end{align}
with
\begin{align}
\beta_{jj'}&=\frac{-3}{2R_{jj'}(R_{jj'}+r_{0,jj'})^4}\,, \\
\gamma_{jj'} &=  \frac{3}{2R_{jj'}^3(R_{jj'}+r_{0,jj'})^4}+ \frac{6}{R_{jj'}^2(R_{jj'}+r_{0,jj'})^5}\,,
\end{align}
and the separation matrices $X_{jj'} = x_j - x_{j'}$, $Y_{jj'} = y_j - y_{j'}$ and $R_{jj'} = |\textbf{r}_j - \textbf{r}_{j'}|$.

We can write $E^{(2)}= \vec\Phi^T \cdot \hat M \cdot  \vec\Phi$, 
with $\vec\Phi = (\epsilon_{x,1},\epsilon_{y,1},\dots \epsilon_{x,N_\text{D}} \epsilon_{y,N_\text{D}})$.  
Now, we can diagonalize $\vec M$ to obtain the eigenvalues $\lambda_\nu$, which provide the excitation frequencies of the droplet array, 
$\Omega_\nu=\sqrt{2\lambda_\nu}$. Note that this is an expansion around the equilibrium positions only, and not a perturbation of the individual droplet shape or atom number, so other shape excitations, such as droplet breathing modes, will not be captured by this method. Some example excitations are shown in Figs.~\ref{fig:4}(b2-b5, c1-c3), where the arrow indicates the vector between $\RR_j$ and $\rr_j$ for each droplet $j$.

\end{document}